\documentclass[aps,prd,twocolumn,floatfix,superscriptaddress]
{revtex4}
\usepackage{graphicx,amssymb,url}

\newcommand{\be}{\begin{equation}}
\newcommand{\ee}{\end{equation}}
\newcommand{\ba}{\begin{eqnarray}}
\newcommand{\ea}{\end{eqnarray}}

\def\lsim{\raise0.3ex\hbox{$\;<$\kern-0.75em\raise-1.1ex\hbox{$\sim\;$}}}
\def\gsim{\raise0.3ex\hbox{$\;>$\kern-0.75em\raise-1.1ex\hbox{$\sim\;$}}}

\def\theta{\vartheta}
\def\ap{\approx}

\def\rp{{\cal F}_{\bar p}/{\cal F}_{\bar p+p}}
\def\re{{\cal F}_{e^+}/{\cal F}_{e^++e^-}}

\begin{document}

\title{Antimatter spectra from a time-dependent modeling of supernova remnants}

\author{M.~Kachelrie\ss}
\affiliation{Institutt for fysikk, NTNU, Trondheim, Norway}

\author{S.~Ostapchenko}
\affiliation{Institutt for fysikk, NTNU, Trondheim, Norway}
\affiliation{D.~V.~Skobeltsyn Institute of Nuclear Physics,
 Moscow State University, Russia}

\author{R.~Tom\`as}
\affiliation{II. Institut f\"ur Theoretische Physik,
    Universit\"at Hamburg, Germany}

\date{April 8, 2010}

\begin{abstract}
We calculate the energy spectra of cosmic rays (CR) and their secondaries 
produced in a supernova remnant (SNR), taking
into account the time-dependence of the SNR shock. We model the trajectories 
of charged particles as a random walk with a prescribed diffusion coefficient, 
accelerating the particles at each shock crossing. Secondary production 
by CRs colliding with gas is included as a Monte Carlo process. We find 
that SNRs produce less antimatter than suggested previously:
The positron/electron ratio $\re$ and the antiproton/proton ratio $\rp$
are a few percent  and few $\times 10^{-5}$, respectively. Both ratios do not 
rise with energy. 
\end{abstract}


\maketitle

\vskip0.5cm
{\em Introduction---}%
Measurements of the antimatter fraction of cosmic rays (CR) provide
not only insight into CR physics itself~\cite{galprop}, as e.g.\ 
their propagation in the galaxy, but are also valuable
probes for cosmology and particle physics. In particular, the
annihilation of dark matter (DM) leads to an equal injection
rate of matter and antimatter particles into the Galaxy, while the CR flux 
from astrophysical sources is matter-dominated. A possible
way to detect DM is therefore to estimate carefully the expected
antimatter fluxes from astrophysical sources and to search then for any
excess~\cite{DM}.

The PAMELA collaboration presented recently results of their measurement 
of the positron fraction in CRs, which is rising rapidly from
10 to 100\,GeV~\cite{PAMELA}. 
At the same time, the antiproton ratio measured by PAMELA declines
above 10\,GeV~\cite{Adriani:2008zq}, consistent with expectations.
The conventional estimate for antimatter fluxes from 
astrophysical sources uses as only production mechanism of antimatter
the scattering of CRs on interstellar gas~\cite{galprop}. As discussed
e.g.\ in Ref.~\cite{Serpico:2008te}, the energy dependence of the
Galactic diffusion coefficient, $D\propto E^\delta$ with 
$\delta=0.5-0.6$, is inconsistent with an increase of the antimatter fraction
with energy. By contrast, the spectral shape of fragmentation functions 
leads quite naturally to such a rise in the case of DM annihilations.

The DM interpretation of the PAMELA excess faces however several 
difficulties~\cite{DM}: First, the required rate of positron production is 
larger than expected for a stable thermal relic. As a consequence, either
the annihilation rate has to be enhanced
, or
the DM particle should be unstable with the appropriate life-time.
Second, in gauge boson or quark fragmentation, positron, antiproton and
photon production are tied together and thus one has to postulate a DM particle
annihilating only into electrons and muons. More importantly, assuming
antimatter production by diffusing CRs as the only astrophysical source for
antimatter falls short: Since electrons lose fast energy, the high-energy
part of the $e^-+e^+$ spectrum should be dominated by local sources as 
nearby pulsars, as pointed out already 20~years ago~\cite{pulsar}. Moreover, 
electromagnetic pair cascades in pulsars 
result naturally in a large positron fraction together with a ``standard'' 
antiproton flux.

More recently, supernova remnants (SNR) were put forward as an alternative 
astrophysical explanation for a rising positron fraction~\cite{B09}: 
Positrons created as secondaries of hadronic interactions in the shock 
vicinity participate in the acceleration process and, according to
Ref.~\cite{B09},  should thus have a flatter energy spectrum than primary 
electrons.  It was estimated 
that the resulting positron fraction can explain the PAMELA excess and 
rise up to 50\% at higher energies~\cite{B09}, while subsequently a 
similar mechanism for antiprotons was 
suggested in Ref.~\cite{pp}. Since 
shock acceleration in SNR is 
expected to be the main source for  Galactic CRs~\cite{SNR}, such
a scenario has also important consequences for the interpretations of CR data
as, e.g., the boron-to-carbon ratio~\cite{BC}.

The present work examines the production of secondary $\bar p$ and $e^+$
in SNRs, improving on previous studies~\cite{B09,pp,sarkar} in two respects:
First, we use a Monte Carlo (MC)
approach calculating the trajectory of each particle individually in
a random walk picture. This makes it easy to include interactions and
the production of secondaries. Second, our approach allows 
us to include the time (and spatial) dependence of relevant parameters
describing the evolution of a SNR as, e.g., the shock radius and its 
velocity, the magnetic field
or the CR injection rate  and to test their influence on the CR spectra.
We should also stress what are {\em not\/} the aims of the present 
work: We do address neither the problem of acceleration from a
microscopic point of view nor consider any feedback of CRs on
the shock or the magnetic field. Although the latter processes are 
important to obtain accurate CR escape fluxes, we shall show that our
simplified treatment leads to an upper limit on the secondary fluxes.

{\em Simulation procedure---}%
Shocks around SNRs are supposed to be collisionless, with charged particle
scattering mainly on inhomogeneities of the turbulent magnetic field. We model 
such trajectories by a random walk in three dimensions with step size $l_0(E)$ 
determined by an energy-dependent diffusion coefficient $D$. Diffusion close 
to the shock is usually assumed to proceed in the Bohm regime with the
mean free path $l_0$ proportional to the Larmor radius $R_L$. Thus 
$D(E)=cl_0/3 = c^2 p/(3ef_BB)$,
%
%
where $f_B$ denotes the ratio of the energy density in the turbulent and 
in the total magnetic field.
We neglect the coupling between CRs and the turbulent magnetic field,
assuming that a layer with Bohm diffusion extends far enough into the 
up-stream region. For a constant magnetic field, 
CRs do not escape but are confined in the SNR,
corresponding to an ``age-limited'' scenario for the CR flux from SNRs.

We describe the evolution of the shock in the rest frame of the SNR. 
Then the (yet unshocked) up-stream region is at rest, $v_1=0$, and has 
the density of the surrounding interstellar medium (ISM), 
$\rho_1=\rho_{\rm ISM}$. Assuming a strong shock 
with Mach number ${\cal M}\gg 1$, the shocked down-stream region 
flows with the velocity $v_2=3v_{\rm sh}/R$ and has the density $\rho_2=R\rho_1$.
Here, $R$ denotes the compression ratio $R=(\gamma+1)/(\gamma-1)=4$ 
for a mono-atomic gas with $\gamma=5/3$. We account for the flow,  
adding in the down-stream
region on top of the random walk an ordered movement of the particle
with velocity $v_2$ that is directed radially outwards. Thus a particle
trajectory evolves during the time step $\Delta t=l_0/c$ as
\be
 {\bf x}(t+\Delta t) = 
 {\bf x}(t)+ v_2\, \Delta t \,\theta(r_{\rm sh}-r)\, {\bf e}_r +{\bf l}_0 \,,
\ee
where ${\bf l}_0$ denotes a random step, $r_{\rm sh}$ the position of the
shock, and $\theta(x)$ the step function.

Crossing the shock front, particles are accelerated. We neglect that the 
relative energy gain $\xi=(E_{k+1}-E_k)/E_k$ per cycle $k$ depends on the angle 
of the  trajectory to the shock front, and use for simplicity that on average 
for a non-relativistic shock
%
 $\xi=\frac{4}{3c}(v_2-v_1)=v_{\rm sh}/c$.
%
For the position $r_{\rm sh}$ and the velocity $v_{\rm sh}$ of the
SNR shock we use the  $n=0$ case of the analytical solutions 
derived in Ref.~\cite{TM99}. These solutions connect smoothly
the ejecta-dominated phase with free expansion $r_{\rm sh}\propto t$
and the Sedov-Taylor stage $r_{\rm sh}\propto t^{2/5}$. The acceleration of 
CRs is assumed to cease after the transition to the radiative
phase at the time $t_{\max}$.

As the injected particles diffuse, electrons lose energy via synchrotron 
radiation and inverse Compton scattering, while protons 
can scatter on gas of the ISM producing secondaries
that include antiprotons and positrons. Cross sections and the final state
of pp-interactions are simulated using QGSJET-II~\cite{qgsjet}, 
while we use SIBYLL~2.1~\cite{sibyll1.7} for decays of unstable particles.

The last ingredient for our simulation procedure is an injection model. 
To ease the comparison with the  results of \cite{B09},
we fix the electron/proton ratio $K_{ep}$ at injection to 
$K_{ep}=7\times 10^{-3}$. As injection energy we use $E_0=10$\,GeV. 
In the first model used, the injection rate 
\be 
 \dot N \propto 
 r_{\rm sh}^2 \,v_{\rm sh}^\alpha \,\delta(E-E_0)\,\delta(r-r_{\rm sh}) 
\ee
is proportional to the volume swept out per time by the shock, i.e.\ 
$\alpha=1$ (thermal leakage model~\cite{in1}).
In the second model, the injection rate $\dot N$ is 
proportional to the CR pressure~\cite{in2} and $\alpha=3$.
In the case of model 2, the fraction of particles injected very early
is significantly larger than in model 1.

We use the following parameters to describe the SNR: We choose the
injected mass as $M_{\rm ej} = 4M_\odot$, the mechanical explosion energy 
as $E_{\rm snr} = 5\times 10^{51}$\,erg, and the density of the ISM as    
$n_{\rm ISM} = 2$\,cm$^{-3}$. The end of the Sedov-Taylor phase follows
then as $t_{\max}= 13.000$\,yr~\cite{TM99}.  For the
magnetic field  we use $B = 1\mu$G and $f_B=1$ to ease the comparison with 
the stationary approach of Refs.~\cite{B09,pp,sarkar}.

\begin{figure}
\includegraphics[width=\linewidth]{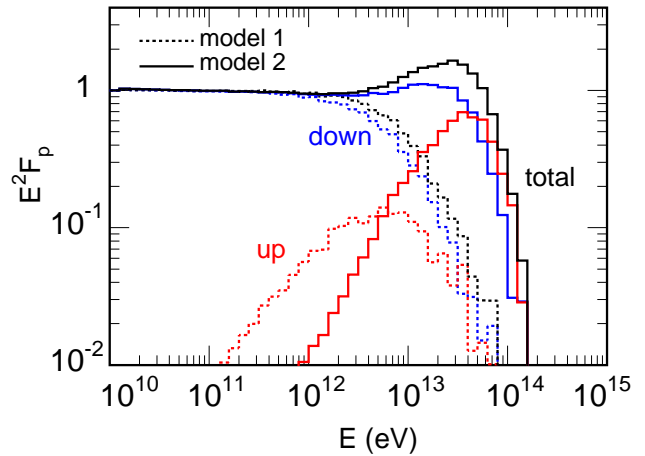}
\caption{
Proton spectra (black) as function of energy for two different injection models
and $f_BB=1\mu$G. 
Additionally the contribution of protons staying until $t_{\rm max}$ in the 
up-stream (red) and of protons in down-stream 
region (blue) are shown.
\label{fig:CR}}
\end{figure}

{\em Numerical results---}%
In Fig.~\ref{fig:CR}, we show the energy dependence of the proton spectra
in model 1  and 2. 
Additionally to the total spectra, the contribution
of protons staying at $t_{\rm max}$ in the up-stream region is shown
in red, while the spectra of protons advected down-stream are
shown in blue.  The spectra of electrons are not shown, 
since they have the same shape as the proton spectra apart from a 
somewhat lower cutoff energy. 
While the total energy spectra  at low energies agree well 
with a $1/E^2$ power-law, changing the injection model leads to large
differences at high energies. The strong dependence of the
spectra close to $E_{\max}$ on the injection model is expected, since 
this part is sensitive to how many particles are injected early, 
when shock acceleration is most effective.

\begin{figure}
\includegraphics[width=\linewidth]{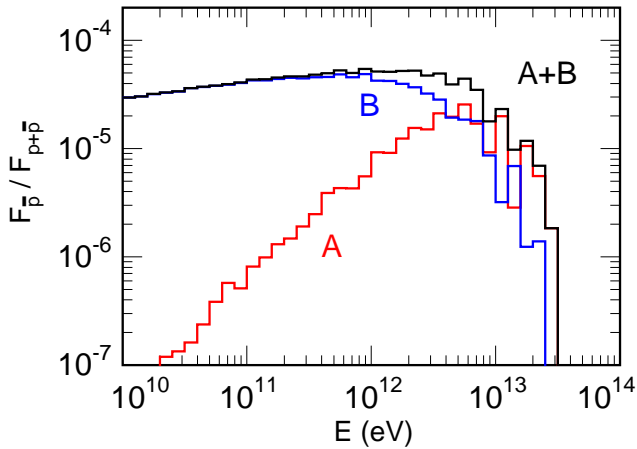}
\caption{
The total flux of antiprotons together with the contribution
A and B in model 2 as function of energy.
\label{fig:rp}}
\end{figure}

\begin{figure}
\includegraphics[width=\linewidth]{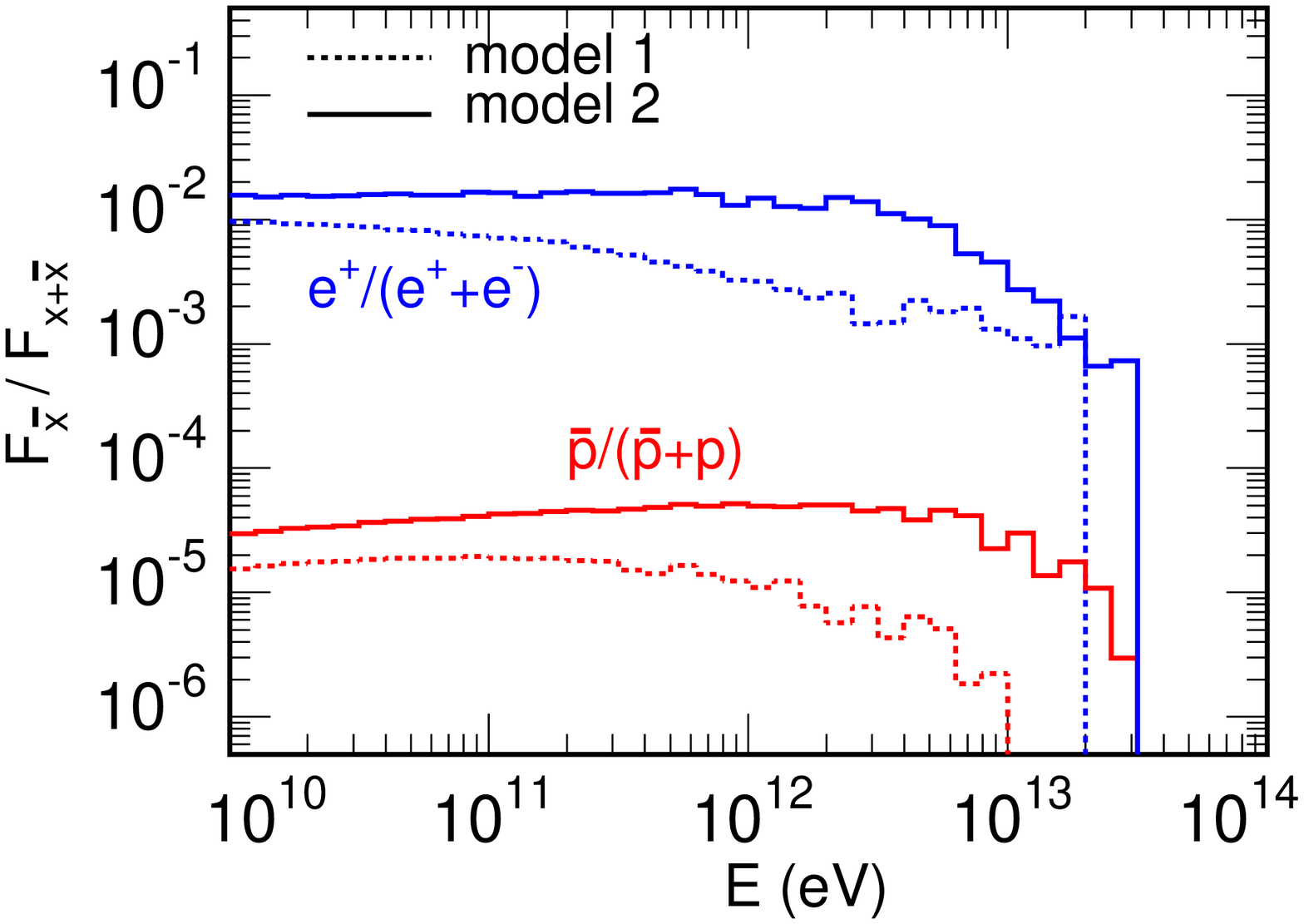}
\caption{
The positron ratio $\re$ (blue) and
antiproton ratio $\rp$ (red) in model 1 (dotted) and 2 (solid).
\label{fig:re}}
\end{figure}

%
We switch now to the produced secondaries and show
in Fig.~\ref{fig:rp} for injection model~2 and $f_BB=1\mu$G 
the antiproton flux split into a
part produced in the acceleration zone (A) and a part produced
in the inner part of the SNR (B). More exactly, we define the contribution 
A as all secondaries that crossed at least once the shock. This contribution 
increases fast, since the time $t_{\rm acc}$ primary protons stay in the 
acceleration zone and can interact increases as 
$t_{\rm acc}\propto D(E)\propto E$. Hence for the  {\em relative\/} rise of $A$ 
not the acceleration of secondaries but of primary protons is
important. For example, the component A for neutral secondaries as
 e.g.\ photons, defined formally as all the particles   produced up-stream,
rises with energy in the same way as the one of antiprotons. 
Since the inelasticity, i.e.\ the energy fraction
$\langle z_{\bar p}\rangle\ap 0.02$
%
%
 transferred to {\em all\/} antiprotons is practically 
constant in the relevant energy range, 
$E_0/\langle z_{\bar p}\rangle\gsim 10^{11}$\,eV, 
its  does not influence the shape of the antiproton flux
\footnote{Note also that, in 
contrast to the assumptions of Refs.~\cite{B09,pp,sarkar},
the average energy fraction per {\em single\/} antiproton (or positron)
$\xi_i=\langle z_i\rangle/\langle n_i\rangle$ decreases strongly with energy,
since the multiplicity $n_i$ in pp interaction increases fast.}.
At $E_{\rm b}\ap 2\times 10^{12}$\,eV, the increase of contribution A 
stops, the total flux retains its approximate $E^{-2}$ slope  
and stays small in contrast to the result of Ref.~\cite{pp}.

How can we understand this behavior and the  maximal value of 
$\rp$? We may assume in a gedankenexperiment that in each 
$pp\to \bar p+X$ interaction the most energetic antiproton 
carries away all the energy, $E_{\bar p}\ap \langle z\rangle E_{p}$ with 
$\langle z\rangle\ap 1$. Then interactions just convert part of the $p$
into a $\bar p$ flux. But since $p$ and $\bar p$ diffuse and are accelerated 
in the same way,  the total $\bar p$ flux is not affected if the $\bar p$
or the parent proton is accelerated. Hence the total flux 
of antiprotons produced in the acceleration zone and inside the SNR,
i.e.\ the sum of A and B, should be simply the proton flux scaled down 
by a constant factor. In particular, the secondary flux of the species $i$ 
is bounded by the proton interaction depth $\tau$ and the 
(spectrally averaged) energy fraction $\langle z_i\rangle$ transferred to $i$.
The maximal conversion rate during the life-time of a SNR is 
with $\sigma_{\rm inel}=30$\,mb as inelastic pp cross section at
100\,GeV given by
$\tau = c\,t_{\max}\,R\,n_{\rm ISM}\,\sigma_{\rm inel}\ap 3\times 10^{-3}$. 
The mean energy fraction of antiprotons (plus antineutrons) is 
$\langle z_{\bar p}\rangle\ap 0.02$, so we may expect a maximal 
ratio  of $\rp\sim \langle z_{\bar p}\rangle\,\tau
 \sim 6\times 10^{-5}$. The obtained $\rp$ ratio shown in  
Fig.~\ref{fig:rp} is indeed close to this estimate.

The relative size of the partial contributions A and B can be understood
considering the relation between the time $t_{\rm acc}$ spent by protons in 
A, their final energy $E\propto t_{\rm acc}$ and thus the interaction
depth $\tau_A$ in A as function of energy, $\tau_A\propto t_{\rm acc}\propto E$.
In particular, it takes all the  life-time $t_{\max}$ of the  SNR to 
accelerate protons to the highest energies, cf. the up-stream component
in Fig.~1. For the component B,
the optical depth $\tau_B$ of the parent proton is  
$\tau_B\propto  (t_{\max}-t_{\rm acc})$,
which  explains why the two components
A and B sum up to a flat spectrum. Note that the relative normalization
of component A and B in the stationary approach of Refs.~\cite{B09,pp,sarkar}
has to be imposed by hand, since B is formally infinite.

The same discussion applies to the case of positrons, with the sole 
exception that the primary electron flux is scaled down by the factor 
$K_{ep}$ and that the energy fraction transferred to positrons is 
$\langle z_{e^+}\rangle\ap 0.05$. The results of our simulation are shown
in Fig.~\ref{fig:re}, confirming with the maximal value of 
$\re\lsim {\rm few}\;\%$  this simple picture. Note that while
above $\sim 100$\,GeV the ratio $\re$ from SNR starts to be larger than
the conventional prediction using only secondary production on the ISM,
it cannot explain the rise to $\re\ap 10\%$ at 10\,GeV in the PAMELA
data~\cite{PAMELA}.


\begin{figure}
\includegraphics[width=\linewidth]{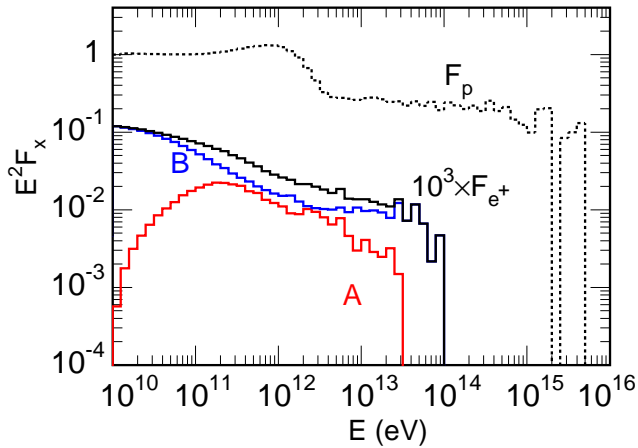}
\caption{
Spectra of cosmic rays protons and positrons (scaled up by a factor 1000)
together with the partial contributions
A and B in model 2  
for a time-dependent diffusion coefficient.
\label{fig:t}}
\end{figure}

Up to now, we have discussed only our numerical results for constant 
$f_BB=1\mu$G and one may wonder if a ``better'' choice of parameters can 
increase the antimatter fluxes. In particular, the analytical formula
of Ref.~\cite{B09,pp,sarkar} seem to imply that the contribution A increases 
for weaker diffusion, i.e.\ larger $D$. However, the term $D/v_1^2$
regulating the importance of A limits also via $t_{\rm acc}\propto D/v_1^2$
the maximal proton energy. Using a constant value $f_B=1/20$ as in 
Ref.~\cite{B09,pp,sarkar}
thus reduces $E_{\max}$  by the same factor. In the stationary
approach,  however, $E_{\max}$ is an external parameter and by increasing 
$E_{\max}$ relative to its natural value given by $t_{\rm acc}=t_{\max}$
one enlarges the relative contribution of A. This approach has been
justified as a method to account in an effective way for amplification
and damping of the magnetic field.

In our time-dependent approach, we test this suggestion considering a
time-dependent magnetic field: assuming that non-linear effects 
amplify magnetic fields in the early phase~\cite{B}, with $f_BB = 100\mu$G before
 the transition to the  Sedov-Taylor phase at $t_\ast=240$\,yr, 
 while in the late stage magnetic fields are damped, $f_BB = 1/20\mu$G at
 $t>t_\ast$.
In Fig.~\ref{fig:t}, we show for this case the proton and positron
spectra using the 
injection model 2. Protons that were injected early are accelerated up to 
few$\;\times 10^{15}$\,eV, while the bulk of CRs injected when the tubulent
 magnetic field is damped has a cutoff  around $10^{12}$\,eV.
The contribution A to the positron
flux saturates at $E\sim {\rm few}\;\times 10^{11}$\,eV, i.e.\ at
the energy expected for  $f_BB = 1/20\mu$G. In contrast to Fig.~\ref{fig:rp},
the component B dominates now the high-energy end of the positron flux:
Most CRs escape from the acceleration zone during the transition 
$f_BB/\mu{\rm G}=100\to 1/20$ and those advected downward contribute
the new high-energy extension of component B, while those escaping
up-stream contribute to A but are not longer accelerated.


Finally, we stress that the splitting between contribution A and B 
is artificial and depends as well as the 
total flux on the definition of the escape flux: If the diffusion 
coefficient drops above a certain energy and/or outside a sufficiently small
radius $r_{\rm sh}+\delta r$ to a value close to the one typical for the 
Galaxy, then CRs can escape up-stream instead of being confined down-stream.
Clearly, this effect reduces the contribution B. On the
other hand, the bounds  $\re\lsim {\rm few}\;\%$  and   
$\rp\lsim   6\times 10^{-5}$ will become stronger, since also the time for 
interactions in the acceleration zone will be shortened.
Since our maximal values of $\rp$ and $\re$ depend only on $t_{\max}$, which is
lower in escape-limited than in age-limited models, we conclude that the 
contribution of SNR to the observed antimatter in CRs does not
lead to rising antimatter fractions and is smaller than estimated in
earlier works.

{\em Summary---}%
We calculated the energy spectra of CRs and their secondaries produced 
in a supernova remnant using a simple random walk picture. In contrast
to a previous prediction that the positron fraction $\re$ can rise up to 
40\%--50\% for $K_{ep}=7\times 10^{-3}$, we found that the ratio levels 
off at a few percent. This
value corresponds to the expectation combining the interaction depth 
$\tau \ap 3\times 10^{-3}$ of a proton during the life-time of a SNR with 
the energy fraction $z\sim 0.05$ transferred to positrons. Similarly,
the antiproton ratio $\rp$ does not rise beyond few$\times 10^{-5}$. 
Our results suggest that antimatter production in SNRs cannot explain 
the rise of the positron fraction observed by PAMELA. Since a rising
antiproton fraction is neither expected from CR interactions with the ISM 
nor from pulsars, such a measurement could be used as a signature 
for dark matter.

{\em Acknowledgments---}%
We thank Pasquale Blasi and Pasquale Serpico for critical comments.
S.O.\  acknowledges a Marie Curie IEF fellowship.
This work was partially supported by the Deutsche Forschungsgemeinschaft 
(SFB 676) and Norsk Forskningsradet  (Romforskning).


\end{document}